\documentclass[12pt,preprint]{aastex}

\def\gsim{\raise0.3ex\hbox{$>$}\kern-0.75em{\lower0.65ex\hbox{$\sim$}}}
\newcommand{\om}{\Omega_{\rm M}}
\newcommand{\ola}{\Omega_{\rm\Lambda}}
\usepackage{epsfig}

\begin{document}

\title{Probing galaxy density profiles with future supernova surveys}

\author{Edvard M\"ortsell\footnote{edvard@astro.su.se}}
\affil{Department of Astronomy, Stockholm University, SE-106 91 Stockholm, Sweden}

\author{H{\aa}kon Dahle\footnote{hdahle@astro.uio.no}}

\affil{Institute of Theoretical Astrophysics, University of Oslo, 
P.O.\ Box 1029, Blindern, N-0315 Oslo, Norway}

\author{Steen Hannestad\footnote{hannestad@fysik.sdu.dk}}
        
\affil{Physics Department, University of Southern Denmark, Campusvej 55, DK-5230 Odense M, Denmark \\
and \\ NORDITA, Blegdamsvej 17, DK-2100 Copenhagen, Denmark}

\begin{abstract}
In this paper we discuss the possibility to measure the Hubble
parameter and the slope of galaxy density profiles using future
supernova data. With future supernova surveys such as SNAP, large
numbers of core collapse supernovae will be discovered, a small
fraction of which will be multiply imaged. Measurements of the image
separation, flux-ratio, time-delay and lensing foreground galaxy for
these systems will provide tight constraints on the slope of galactic
halos as well as providing complementary and independent information
to other cosmological tests with respect to the Hubble
parameter. 
\end{abstract}


\section{Introduction}
In recent years, precise measurements of the Cosmic Microwave
Background (CMB), large scale structure (LSS), and distant supernovae
(SNe) have increased our knowledge of the large scale properties of
the universe tremendously. The standard $\Lambda$CDM model of
cosmology has turned out to be able to fit all available data with
only a few free parameters.

However, there are still many parameters related to cosmology which
are not well determined by the present large scale observations. For
instance, in order for the CMB data from the WMAP experiment
(\citealp{Spergel:2003cb,Bennett:2003bz,Kogut:2003et,Hinshaw:2003ex,Verde:2003ey,Peiris:2003ff}) to
provide a stringent limit on the curvature of the universe, a prior on
the Hubble parameter is necessary. This degeneracy can be broken by
adding LSS data from surveys such as the Sloan Digital Sky Survey
(SDSS) \citep{Tegmark:2003uf,Tegmark:2003ud} or the 2dF survey
\citep{2dFGRS}, but other independent methods for measuring the Hubble
parameter precisely are of great interest.

Another problem is that the nature of the
dark matter is at present unknown. The properties of dark matter on
large scales are compatible with a heavy, collisionless species which
clusters gravitationally, cold dark matter (CDM). On the other hand
there are significant problems with understanding some properties of
galaxies in CDM models. Numerical simulations of galaxy formation in
CDM models show that dark matter halos exhibit a universal density
profile with a central slope $\rho \propto r^{-\eta}$, with $\eta
\simeq 1-1.5$
(\citealp{Navarro:1996he,Ghigna:1999sn,Power:2002sw,Fukushige:2003xc,Hayashi:2003sj,Kazantzidis:2004hb,Navarro:2004ew}). 
At large radii the profile steepens
to $\rho \propto r^{-3}$. While the outer slope is consistent with 
current observations, there are problems with fitting the inner slope of
halos.

Observations of rotation curves of many dwarf galaxies, indicating
almost constant density cores, suggest that the inner density profiles
in these systems are much shallower than found in simulations
\citep{deBlok:2001fe,vandenBosch:1999ka,Dutton:2003ib,Simon:2003tf,%
Simon:2003xu}. Another problem is that the number of
subhalos within a Milky Way-sized halo predicted by the
simulations exceed the observed number of satellite galaxies by at
least an order of magnitude.

Many explanations have been offered for this discrepancy. One
possibility is that tidal interactions between baryons and dark matter
erase the central cusps, and that early star formation expels baryons
from small subhalos, rendering them invisible. Whether this purely
astrophysical explanation can work is not clear at present.
Another possibility is that the properties of dark matter are slightly
different from pure CDM. If for instance dark matter has relatively
strong self-interactions or is warm instead of cold, shallow density
profiles and lack of subhalos might be explained. However, it should 
be noted that the simplest models of self-interacting dark matter  
predict large constant-density cores in clusters of galaxies which are 
ruled out by X-ray and gravitational lensing observations 
(Meneghetti et al.\ 2001; Arabadjis, Bautz, \& Garmire 2002; 
Dahle, Hannestad, \& Sommer-Larsen 2003). 

At present the ultra-high resolution simulations of galaxy formation do not include baryons. This in turn means that for regions of galaxies which are baryon dominated the discrepancy might be resolved by a more complete understanding of feed-back from star formation. In dwarf galaxies this
effect is unlikely to be dominant, but for the high-mass systems typically investigated with the method proposed here, the regions probed will be dominated by the baryonic component. Therefore the density profile which we derive for the inner parts of galaxies is likely to be dominated by baryonic physics and can therefore not be directly related to either the pure N-body simulations or to the observations of dark matter dominated dwarf galaxies.

Nevertheless it is of great importance to get a precise observational
determination of the halo density profiles in order to understand
whether the difference between observations and simulations is
generic, or specific to some systems. 

Gravitational lensing has the potential to give information on both
the matter distribution in galaxies as well as the Hubble
parameter. More than 40 years ago, Refsdal outlined how it should be
possible to measure the Hubble parameter and the mass of a galaxy by
measuring the time delay between multiple images of a supernova (SN)
lying far behind and close to the line of sight through the lensing
galaxy
\citep{refsdal}.

The first observation of gravitational lensing in a cosmological
context was made in 1979, when two images of the quasar QSO 0957+651
at $z\sim 1.4$ was observed. For several reasons, it has been
difficult to implement Refsdal's method on this system. Until 1997 it
was not possible to precisely determine a time delay for the system,
and even then the complexity of the lens system (a brightest 
cluster galaxy located close to the center of a galaxy cluster) 
has made estimates of the 
Hubble parameter highly uncertain \citep{keeton}. Studies of this
system have also showed the importance of including effects from
substructure in the lens.

To date, approximately 70 multiply imaged sources have been observed
out of which there are eleven firm time delay measurements
(\citealp{davis,Kochanek:2003};\\
\citealp{york}). There is not yet any agreement on the correct 
value of the Hubble parameter inferred from these measurements due to
differences in the modeling of the lensing galaxies. In general,
modeling the lenses as more concentrated gives larger values of $h$
and vice versa, e.g., Kochanek (2002) obtains $h\sim
0.48^{+0.07}_{-0.04}$ if the lenses have isothermal mass distributions
and $h\sim 0.71\pm 0.06$ if they have constant mass-to-light ratio.

There are basically two different routes in trying to improve on
current results. One is to try to do extremely careful modeling of
single simple lens systems. Following this approach, Wucknitz, Biggs,
\& Browne (2004) find $h = 0.78\pm 0.06$ (2 $\sigma$) based on the
lens system B0218+357, consistent with the local estimate $h = 0.71\pm
0.06$ (1 $\sigma$) from the Hubble Space Telescope (HST) key project
(Mould et al.\ 2000; Freedman et al.\ 2001). Another possibility is to
combine results from a larger number of, perhaps less well
constrained, systems. In this paper, we will investigate the latter
method.

In order to avoid any systematic bias due to selection effects, it is
important to have a well-defined statistical sample. The currently
largest survey of strong lensing events is the recently completed
Cosmic Lens All-Sky Survey. A subsample of 8958 flat-spectrum radio
sources of which 13 have multiple images constitutes a well-defined
statistical sample and has been used to constrain cosmological
parameters as well as galaxy properties
\citep{davis,chae}.
  
Future surveys that will improve statistics and also include multiply
imaged SNe for which time delay measurements can be obtained at high
accuracy due to the characteristic light curves include Pan-STARRS
\footnote{{\tt http://www.pan-starrs.org}} 
 and the Supernova Acceleration Probe (SNAP)
\footnote{SNAP Science Proposal, available at {\tt http://snap.lbl.gov}}.

In this paper we investigate how future SNAP data of multiply imaged
core-collapse SNe can be used to measure the slope of galactic halos
and the Hubble parameter (see also 
\citealp{2001ApJ...556L..71H,2003ApJ...583..584O}).
This paper extends and generalizes earlier work 
\\ \citep{Goobar:2002a}
where similar data was used to constrain $h$ and $\om$ using simpler
lens models.  A recent work by Oguri \& Kawano (2003) have considered
a similar use of quadruply lensed Type Ia SNe events. The possiblity
of using SNe that are multiply lensed by rich 
galaxy clusters to constrain $h$ 
has been considered by Bolton \& Burles (2003). The complementarity of
strong lensing to other probes of the expansion rate is discussed by
Linder (2004).

Finally, it should be noted that while the method proposed here is
very sensitive to the inner density profile of galaxies, this density
profile does not necessarily reflect the nature of pure dark matter
halos. Lensing mainly occurs in relatively massive galaxies in which
baryonic material is a dominant component at small radii. This in turn
means that there can be a substantial modification of the inner
density profile due to cooling and infall (\citealp{kocwhite}).

In Sec.~\ref{sec:method}, we describe the lens model used in the
simulations and analysis and in Sec.~\ref{sec:simdata} we describe our
numerical simulations. Errors are discussed in Sec.~\ref{sec:errors}
and our results are presented in Sec.~\ref{sec:results}. Finally,
Sec.~\ref{sec:summary} contains a summary of our results.

\section{Method}\label{sec:method}

\subsection{Lens model}
\begin{figure}[htb]
\plotone{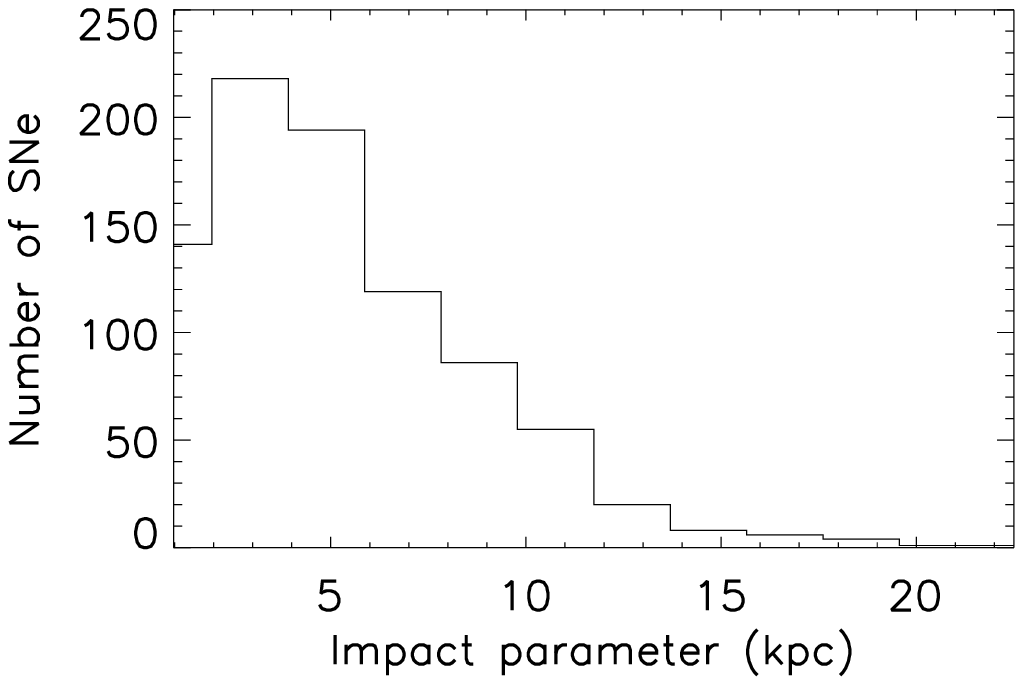}
  \caption{Distribution of impact parameters for primary images for
  the lens systems used in the analysis in this paper.}
\label{fig:impact}
\end{figure}

For most lensing purposes, galaxy matter distributions can be
described by the projected Newtonian gravitational potential
\begin{equation}
    \Psi =\frac{2D_lD_{ls}}{c^2D_s\xi_0^2}\int\Phi dl.
\end{equation}
We assume that lens systems are characterized by a simple power-law density profile 
\begin{equation}
\rho(r) \propto r^{-\eta}.
\end{equation}
This is not a good approximation at all radii since the slope is
expected to change with radius. In practice however, almost all images
of multiply lensed SNe are at a limited range of small $r$ (see
Fig.~\ref{fig:impact}) and thus only the inner slope is probed and the
single power-law density profile is an excellent approximation.

For the power-law density profile, the lenses can be described by the
projected potential
\begin{equation}\label{eq:psi}
    \Psi =\frac{x_E^{\eta -1}}{3-\eta}x^{3-\eta}
\end{equation}
where $x$ is the impact parameter (in arbitrary units, $\xi_0$) and
$x_E$ is the Einstein radius.

\subsection{Time delay}
The time delay for a gravitationally lensed image as compared to an
undeflected image is in the general case given by (Schneider et al. 1992)
\begin{equation}\label{eq:T}
    T=\frac{\xi_0^2D_s}{D_lD_{ls}}(1+z_l)
    \left[\frac{({\bf x} -{\bf y})^2}{2}-\Psi \right],
\end{equation}
where $\xi_0$ is the (arbitrary) scale length, the $D$'s are angular
diameter distances and $y$ is the source position. The time delay
between two lensed images is given by $\Delta t=T_2-T_1$.  For an
isothermal lens, we have $\eta =2$ and
\begin{equation}
    \Delta t_{\rm SIS}=\frac{\xi_0^2}{2}\frac{D_s}{D_lD_{ls}}(1+z_l)
    (x_1^2-x_2^2).
\end{equation}
Putting $\xi_0=D_l$, i.e., denoting positions in terms of angles we get
\begin{equation}
    \Delta t_{\rm SIS}=\frac{1}{2}\frac{D_lD_s}{D_{ls}}(1+z_l)
    (\theta_1^2-\theta_2^2).
\end{equation}
In the general case of $\eta\ne 2$, we can rewrite Eq.~(\ref{eq:T}) in
terms of
$q\equiv\Delta\theta/\hspace{-.1cm}<\hspace{-.1cm}\theta\hspace{-.1cm}>$
where $\Delta\theta =\theta_1-\theta_2$ and
$<\hspace{-.1cm}\theta\hspace{-.1cm}>=(\theta_1+\theta_2)/2$ and
Taylor expand to get [after some tedious calculations; see also 
Kochanek \& Schechter (2003) and Chang \& Refsdal (1977) for an earlier 
approximation]
\begin{eqnarray}\label{eq:delt}
    \Delta t&=&\frac{(\eta -1)}{2}\frac{D_lD_s}{D_{ls}}(1+z_l)
    (\theta_1^2-\theta_2^2)\left[1-\frac{(2-\eta)^2}{12}
    q^2+\mathcal{O}
    (q^3)\right]\nonumber\\
    &\simeq&(\eta -1)\Delta t_{\rm SIS}\left[1-\frac{(2-\eta)^2}{12}
    q^2\right].
\end{eqnarray}
We see that we have an almost perfect degeneracy between $\eta -1$ and
$h$ (which comes in through the distances). This degeneracy can be
broken by including data from the observed flux ratio.

\subsection{Flux ratio}
We denote the flux ratio $r\equiv \mu_1/|\mu_2|$. The magnification
$\mu$ for a source at position $y$ observed at position $x$ is given
by
\begin{equation}
    |\mu|=\left|\frac{x}{y}\frac{dx}{dy}\right|.
\end{equation}
Computing the lensing angles through $\alpha
=\nabla\Psi$ and using the lens equation
\begin{equation}
    y = x_1-\alpha_1=x_2- \alpha_2\, ,
\end{equation}
we can express the flux ratio as
\begin{eqnarray}\label{eq:r}
    r&=&\left|\frac{x_1}{x_2}\right|\left|
    \frac{1-(2-\eta)x_2^{1-\eta}(x_1+x_2)/(x_1^{2-\eta}+x_2^{2-\eta})}
    {1-(2-\eta)x_1^{1-\eta}(x_1+x_2)/(x_1^{2-\eta}+x_2^{2-\eta})}\right|\nonumber\\
    &=&r_{\rm SIS}\left|
    \frac{1-(2-\eta)x_2^{1-\eta}(x_1+x_2)/(x_1^{2-\eta}+x_2^{2-\eta})}
    {1-(2-\eta)x_1^{1-\eta}(x_1+x_2)/(x_1^{2-\eta}+x_2^{2-\eta})}\right|.
\end{eqnarray}
Expanding this to second order in $q$, we obtain
\begin{equation}\label{eq:rq}
r \simeq 1+(\eta-1)q + \frac{1}{2}(\eta-1)^2 q^2  + {\cal O}(q^3).
\end{equation}

Eqns.~(\ref{eq:delt}) and (\ref{eq:rq}) are excellent approximations
for small $q$ and/or $\eta\sim 2$. In order to show the general
parameter dependencies, we generalize in the following the expanded
expressions to include also the effects from ellipticities and
external shear. However, in the subsequent $\chi^2$-analysis, we use
the full expressions for the time delay and the flux ratio to avoid
any bias due to the Taylor expansion\footnote{Since $\eta\sim 2$ in
our simulations, the use of the Taylor expanded expressions gives
identical results.}.

\subsection{Ellipticity}

In order to treat possible non-sphericity of the lens systems we
add a quadrupole term to the projected potential, $\Psi$,
\begin{equation}
\Psi = \Psi_0 \left(1+a \cos 2\phi\right),
\end{equation}
where $\Psi_0$ is the spherical potential given in Eq.~(\ref{eq:psi})
and $\phi$ is the angle of the image relative to the quadrupole axis.
Using this potential it is possible to derive an analytic expression
for the time delay, which is, however, quite complicated. If the
system is assumed to have relatively small ellipticity, an assumption
already implicit in the fact that we represent the asphericity of the
lens with a quadruple moment, then the angle of the second image can
be written as a function of the first
\begin{equation}
\phi_2 = \phi_1 + \pi - \delta,
\end{equation}
where $\delta$ is a small parameter (and explicitly 0 for
spherical systems). We expand the time delay in the parameters
$q$ and $\delta$ to find
\begin{equation}
\Delta t \simeq (\eta-1) \Delta t_{\rm SIS}
\left[1-\frac{(\eta-2)^2}{12} q^2 - \frac{\eta-2}{4} \delta^2
\right],
\end{equation}
an expression which is valid to second order in both $q$ and
$\delta$. In the same way it is possible to derive an expression
for the flux ratio of a system with a quadrupole moment. To second
order in $q$ and $\delta$ we find that
\begin{eqnarray}
r & = & 1 + 2 \delta \cot(2 \phi_1) + \frac{3+\cos (4
\phi_1)}{\sin^2(2 \phi_1^2)} \delta^2 \nonumber \\
&& \,\, + (\eta-1)\left[1+ \delta \cot(2\phi_1)\right]q + \frac{1}{2}(\eta-1)^2
q^2 +{\cal O}(q^3,\delta^3, q^2 \delta, q \delta^2) .
\end{eqnarray}
The above expression appears divergent for $\phi_1 \to 0$ because of
the $1/\sin(2\phi_1)$ terms. However, since $\delta \to 0$ when
$\phi_1 \to 0$, the quantity $\delta/\sin(2\phi_1)$ remains finite,
and there are no actual divergences.

\subsection{External shear}

Most lens systems are embedded within an external potential which
gives rise to an additional shear contribution. Following Kochanek
\& Schechter (2003) we write the combined potential as
\begin{equation}
\Psi = \Psi_0 \left(1+a \cos 2 \phi\right) + \gamma_{\rm ext} r^2
\cos[2(\phi-\phi_0)],
\end{equation}
where $\phi-\phi_0$ is the angle between the image position and
the quadrupole axis of the external potential. If the external
shear is aligned with the lens system itself then $\phi_0=0$, but
in the general case $\phi_0 \neq 0$.

Since $\gamma$ should be small in order to justify treating the
external potential as a quadrupole (an assumption supported in, e.g.,
\citealp{dalal04}), we derive expressions for $\Delta t$ and $r$ which
are valid to second order in $q$, $\delta$ and $\gamma$ for arbitrary
values of $\phi_0$. The expressions we find are
\begin{equation}\label{eq:fulldelt}
\Delta t \simeq (\eta-1) \Delta t_{\rm SIS}\Big[1-
\frac{(\eta-2)^2}{12}q^2 + \delta \gamma + \frac{1}{4} \delta^2 (\eta-2)\Big] 
\end{equation}
and
\begin{eqnarray}\label{eq:fullr}
r & \simeq & 1 + (\eta -1)q + 2\delta\cot(2\phi_1) +
 \frac{1}{2}\left[q(\eta -1) + 2\delta\cot(2\phi_1)\right]^2 \\ \nonumber && + \frac{\gamma^2}{8(\eta -1)\sin(\phi_1)^2\cos(\phi_1)^2}
    \Bigg[\Big(\eta(10 + \eta)-43\Big)\cos(4\phi_0) + 
    (\eta -1)^2\cos[4(\phi_0 - 2\phi_1)]\\ \nonumber
&&    +2[23 + (\eta -8)\eta - (\eta -3)(\eta -1)
       \cos[4(\phi_0 - \phi_1)] - (\eta^2 -1)
       \cos(4\phi_1)]\Bigg] \\ \nonumber 
&& -\frac{8\gamma\cos(\phi_0)\sin(\phi_0)}{\sin(\phi_1)\cos(\phi_1)} - 
\frac{\gamma}{16(\eta -1)}\Bigg[16\delta (\eta -1)^2\cos(2\phi_0)
     \sin(2\phi_1)    \\ \nonumber
&& + \frac{\sin(2\phi_0)}{\sin(\phi_1)^2\cos(\phi_1)^2}
\Big(- \delta[81 + (\eta -98)\eta]
        \cos(2\phi_1) +  (\eta -1)
       [\delta(\eta -1)\cos(6\phi_1) +
        8q\eta\sin(2\phi_1)]\Big)\Bigg] .
\end{eqnarray}

\section{Simulated data}\label{sec:simdata}
With the proposed SNAP satellite a large number of core collapse
supernovae (CC SNe) will be discovered, of which a small fraction will
be multiply imaged.  We use the SNOC package \citep{snoc} to simulate
a total of $1.1\cdot 10^6$ CC SNe, the predicted number following the
prescriptions in Dahl{\'e}n \& Fransson (1999) for a 20 square degree
field during three years for $z<5$. Lensing effects are investigated
by ray-tracing using SIS halo profiles since lensing statistics are
insensitive to ellipticities and shear \citep{huterer}. We derive a
galaxy mass function by combining the Schechter luminosity
function with a mass-to-luminosity relation for fundamental plane
ellipticals; see Bergstr\"om et al. (2000) for details.

The cosmology used is $h=0.65$, $\om =0.3$ and $\ola =0.7$. Out of the
$1.1\cdot 10^6$ CC SNe, 2613 have multiple images [this is the sample
used in Goobar et al. (2002a)]. Out of these, 857 have either an
I-band or J-band peak brightness $<28.5$ mag for the dimmest image.
In order to avoid contamination from the lens galaxy in the SN
detection, we impose a constraint that the surface brightness of the
lens galaxy has to be fainter than 24 I magnitudes per square
arcsecond. We assume that the lens galaxies follow a de Vaucouleurs
profile \citep{deVauc}
\begin{equation}\label{eq:i}
I(r) = I_{e} \exp \left\{ -7.67 \left[ (r/r_{e})^{1/4} - 1 \right] \right\}
\end{equation}
where $I(r)$ is the surface brightness at radius $r$, $r_e$ is the half
light radius and $I_{e}$ is the surface brightness at
$r_e$. Observationally, the surface brightness at $r_e$ (in B magnitudes
per square arcsecond), $\mu_{Be}$, can be related to the half light radius 
according to \citep{kormendy}
\begin{equation}\label{eq:b0v}
\mu_{{\rm B}e} = 3.02\log (r_e/{\rm 1\,kpc}) +19.74.
\end{equation}
The absolute B-band magnitude, $M_{\rm B}$, can be written
\begin{equation}\label{eq:m_b}
M_{\rm B} = \mu_{{\rm B}e}-5\log (r_e/{\rm 1\,kpc})-39.28.
\end{equation}
We use Eqns.~(\ref{eq:i})-(\ref{eq:m_b}) together with the
Faber-Jackson relation $L\propto v^4$ and a 
K-correction appropriate for early-type galaxies to calculate the I-band 
surface brightness, $\mu_{\rm I}$ at the position of the faintest
image. Here, we have neglected the modest surface brightness evolution of 
elliptical galaxies to $z \sim 1$. 
All but 4 of the 857 SNe satisfy $\mu_{\rm I} > 24$ at this 
position and we conclude that the lens galaxy light does not significantly
affect the SN observations.

Since we assume fairly clean and simple lens systems in our analysis,
we incorporate a quality factor, $f$, giving the fraction of the lens
systems that fulfill our requirements of a single dominant deflector
with not too large ellipticity or external shear. In accordance with
currently observed systems (see, e.g., Kochanek \& Schechter 2003), we
set $f=0.5$ in our analysis. We also perform an extremely conservative
analysis setting $f=0.05$.

Even though we can derive fair lensing statistics using a simple SIS
galaxy profile, it is important to include lens ellipticities and
galaxy environments in the analysis\\ \citep{keetonzabludoff}. 
We modify our simple simulated lensing systems by adding the effects of
ellipticities and shear according to the distributions described in
Sec.~\ref{sec:errors}. Since we expect a scatter in the properties of
individual halos, we have also added a 1 sigma dispersion in the value
of $\eta$ for individual halos of $\Delta\eta =0.2$. However, in order
to constrain the mean slope of galactic halos, we are still fitting
$\eta$ as a global parameter.

\section{Assumed errors}\label{sec:errors}

\paragraph{$\sigma_r$:}

Errors in the flux ratio of the two SN images will be dominated
by the effects of microlensing by compact objects (Schneider \&
Wagoner 1987) and millilensing by cold dark matter (CDM) sub-halos
(Dalal \& Kochanek 2002; Keeton 2003; Mao et al.\ 2004), rather than
flux measurement uncertainties.

In QSO microlensing, a source of essentially constant size is observed
moving across a microlens amplification pattern at a typical
transverse velocity of $\sim 500\, {\rm km\, s}^{-1}$. In the case of
microlensing of a SN, this transverse velocity is dwarfed by
the $\sim 10^{4}\, {\rm km\, s}^{-1}$ expansion velocity of the
SN photosphere. If the center of the SN is close to a
critical line, an amplification of several magnitudes may result from
microlensing (Schneider \& Wagoner 1987). In this case, the shape of
the light curve of the affected SN image is usually strongly
modified, and such cases should be readily identifiable from a comparison of
the light curves of the two SN images.
 
The light curves of QSO 2237+0305 (Huchra's lens) presented by the
OGLE team \\ \citep{woz} indicate a scatter $\sigma_r = 0.4$ caused by
stellar microlensing. When also taking into account a $\sim 20\%$
uncertainty in the image fluxes due to CDM substructure, we arrive at
a conservative estimate of $\sigma_r/r = 0.5$, which was used for our
simulations.

\paragraph{$\sigma_{\theta}$:}

Following Goobar et al. (2002a), we conservatively estimate a
positional uncertainty of $\sigma_{\theta} =0.01''$ for SNAP
data. While millilensing by CDM sub-halos may
affect the flux ratio, its effect on image positions would likely be
negligible, compared to the measurement uncertainty.

\paragraph{$\Delta t$:}

The estimated measurement uncertainty for the time delay is 0.05 days
(Goobar et al.\ 2002a). However, in the presence of microlensing, the
measured time delay may depart significantly from the time delay
predicted by our smooth macro-lens model. In the most extreme cases, a
high-amplification microlensing event may shift the observed maximum
by several days (Schneider \& Wagoner 1987), but as noted above, such
rare events should be identifiable and can thus be removed from the
sample of events. We model the observed SN light curve as the
intrinsic curve superposed on a smooth gradient, with a typical value
similar to the microlens light curves of Wozniak et al.\ (2000). From
this, we estimate an uncertainty $\sigma_{\Delta t}$ of 0.15 days.

\paragraph{$\gamma_{\rm ext}$:}

Using N-body simulations and semi-analytic models of galaxy formation,
Holder \& Schechter (2003) estimate a mean external shear at the
position of lens galaxies of $\gamma_{\rm ext} = 0.058$, with an rms
dispersion of 0.071.
They also find a significant tail towards high values of $\gamma_{\rm
ext}$. We note that the photometric redshift estimates obtained with
SNAP should enable the identification -- and removal -- of a small
subset of lens systems where nearby galaxy groups or clusters produce
large values for the external shear.
 
\paragraph{Quadrupole:}

Detailed models of known gravitational lens systems indicate that the
mass distribution is aligned with the optical light distribution, with
an upper limit $\langle \Delta \phi^2 \rangle^{1/2} < 10 \degr$ on the
dispersion in the angle between the major axes (Kochanek 2002). The
measurement uncertainties in determining the positions of the lensed 
images with
respect to the lens galaxy will be negligible by comparison. In lens
galaxies undergoing major mergers, a temporary misalignment of the
optical and dark matter may occur (Quadri, M{\"o}ller \& Natarajan
2003). Here, we neglect such effects on the assumption that such
merging systems will have morphological and photometric signatures
that enable their identification and removal from our sample.

\paragraph{$\sigma_{z_l}$:}

We adopt $\sigma_{z_l} = 0.001$ from photometric redshift measurements
based on SNAP multi-band photometry (Goobar et al.\ 2002a).

\paragraph{$\Omega_m$:}

We use $\Omega_m = 0.30 \pm 0.04 $ from the Tegmark et al.\ (2003b)
analysis of SDSS and WMAP data. It should be noted that this is a very
conservative estimate of the error. Future CMB measurements from
experiments like Planck, as well as large scale weak lensing surveys
are likely to increase the precision with which $\Omega_m$ can be
measured.

\section{Results}\label{sec:results}

In order to investigate constraints from the time delay measurements,
we perform a $\chi^2$-analysis over a $[h,\eta]$-grid
\begin{equation}
    \chi^2(h,\eta)=[\Delta t_i^{\rm exp}-\Delta t_i^{\rm th}(h,\eta)]
    V_{ij}^{-1}[\Delta t_j^{\rm exp}-\Delta t_j^{\rm th}(h,\eta)],
\end{equation}
where $\Delta t_i^{\rm exp}$ is the experimental time delays
(simulated using the SIS profile and modified to include the effects
from ellipticity, external shear and a scatter in the slope, see
Sec.~\ref{sec:simdata}) and $\Delta t_i^{\rm th}(h,\eta)$ is the
theoretical time delays. $V_{ij}$ is the error (covariance) matrix.
We expect a close to perfect degeneracy between $h$ and $1-\eta$ that
can be broken by including information from the flux ratio that is
sensitive to the slope $\eta$ only.  For constraints using the flux
ratio, we use
\begin{equation}
    \chi^2(h)=[r_i^{\rm exp}-r_i^{\rm th}(\eta)]
    V_{ij}^{-1}[r_j^{\rm exp}-r_j^{\rm th}(\eta)],
\end{equation}
where $r_i^{\rm exp}$ is the experimental (simulated) flux ratio and
$r_i^{\rm th}(\eta)$ is the theoretical flux ratio. 

In order to avoid any bias due to the second order expansion in $q$ in
the time delay and flux ratio, we use the full expression for these
quantities. However, since $\eta\sim 2$ in our simulations, the use of
Eqns.~(\ref{eq:fulldelt}) and (\ref{eq:fullr}) for the theoretical
time delay and flux ratio, respectively, gives close to identical
results.

\begin{figure}[htb]
\plotone{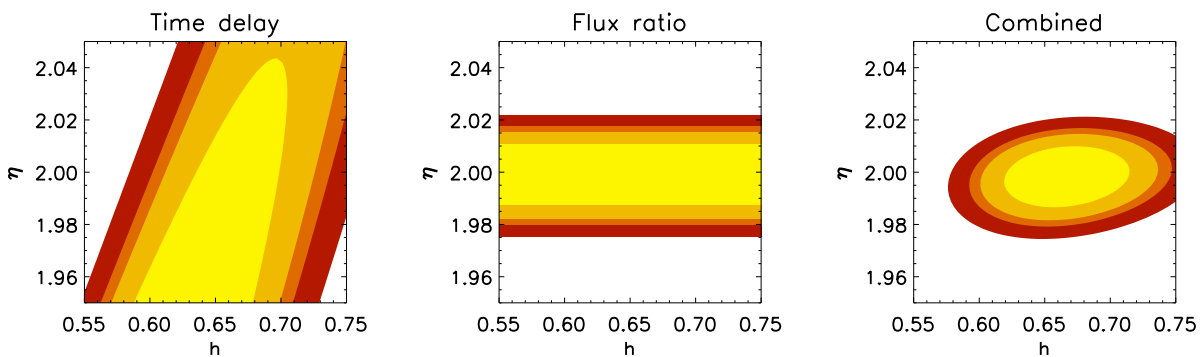}
  \caption{Results from the time delay (left panel), flux ratio
  (middle panel) and the combined results (right panel) using a
  quality factor $f=0.5$ or $\sim 400$ SNe. Contours correspond to
  68.3\,\%, 90\,\%, 95\,\% and 99\,\% confidence levels. }
\label{fig:chi2}
\end{figure}

In Fig.~\ref{fig:chi2}, results from the time delay (left panel), flux
ratio (middle panel) and the combined results (right panel) are
shown. Using a quality factor $f=0.5$, this gives a total of $\sim
400$ SNe for the magnitude cuts described in
Sec.~\ref{sec:simdata}. Contours correspond to 68.3\,\%, 90\,\%,
95\,\% and 99\,\% confidence levels. It is clear that when combining
the results from the time delay and flux ratio measurements, we are
able to make a determination of $h$ within 10\,\% and, perhaps more
interestingly, to determine $\eta$ at the per cent level at 95\,\%
confidence.
 
\begin{figure}[htb]
\plotone{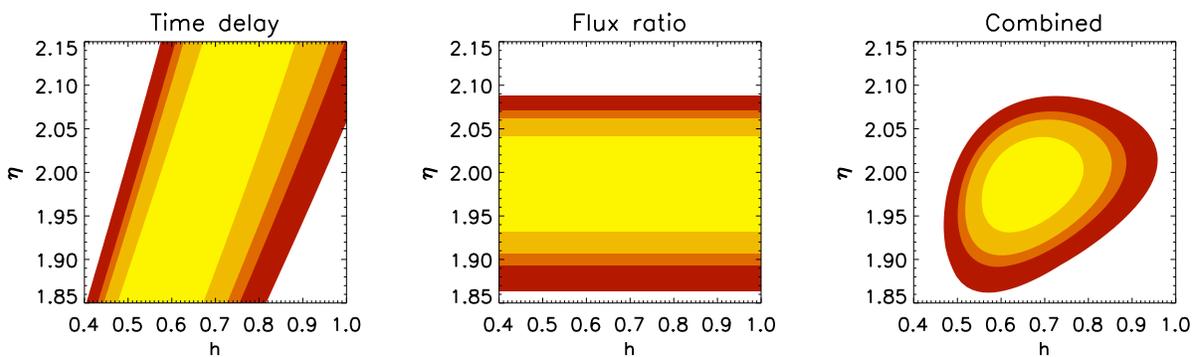}
  \caption{Results from the time delay (left panel), flux ratio
  (middle panel) and the combined results (right panel) using a
  quality factor $f=0.05$ or $\sim 40$ SNe. Contours correspond to
  68.3\,\%, 90\,\%, 95\,\% and 99\,\% confidence levels. Note
  different axis range from Fig.~\ref{fig:chi2}.}
\label{fig:simfrac005}
\end{figure}

In Fig.~\ref{fig:simfrac005}, results using a quality factor $f=0.05$,
giving a total of $\sim 40$ SNe are shown. Even with this drastic
decrease in statistics, we are still able to determine the slope
$\eta$ to an impressive accuracy.

Constraints from the time delay are almost fully determined by the
quality of the time delay measurements whereas the flux ratio
constraints are quite robust to all observational errors. Even when
increasing the error in the flux ratio observation with a factor of 3
(to 150\,\%), we are still able obtain $\eta =2\pm0.06$ at 90\,\%
confidence. Increasing the size of the external shear by a factor of
ten causes a systematic bias in the determination of $\eta$ of
$\eta=2\rightarrow 1.65$. Increasing the value of the dispersion in
the value of $\eta$ for individual halos to $\Delta\eta=0.5$ will
cause a bias in $h$ of $h=0.65\rightarrow 0.55$ but the mean slope
$\eta$ will still be well-determined. Thus, we conclude that our
results are fairly robust even to quite significant changes in the
quality and quantity of the data used in the analysis.

As is evident from Eq.~(\ref{eq:fullr}), the effect on the flux ratio
from varying $\eta$ is strongest for high $q$, i.e, large flux
ratios. For large $q$, we also have large time delays. Since these
systems have smaller fractional time delay errors, we obtain stronger
constraints also from the time delay analysis for large $q$. Thus, we
conclude that systems with large flux ratios and time delays (i.e.,
large $q$) are very important when constraining galaxy density
profiles and the Hubble parameter using strong gravitational lensing.
This is confirmed by dividing the lens system into two classes with
$0<q<1$ and $1<q<2$ where (for an equal number of lens systems) the
class with high $q$ gives confidence contours a factor of $\sim 2$
smaller in the $h$-direction and a factor of $>5$ smaller in the
$\eta$-direction compared with the low-$q$ systems.

In practice, it should be possible to obtain additional constraints 
for a significant fraction of the observed lens systems, since some of 
the SN host galaxies will be lensed into multiple, resolved images that 
may be detectable in deep combined frames from the survey. We make no 
attempt here to model the possible improvements in the parameter 
determination resulting from such additional constraints, but note that 
our results here correspond to a pessimistic scenario where no 
multiply lensed SN host galaxies are detectable. 

\section{Summary}\label{sec:summary}

We have analyzed how the measurement of strongly lensed SNe by future
SN surveys will allow for a very precise determination of statistical
properties of the matter distribution in galaxies, as well as the
Hubble parameter.

Understanding the properties of dark matter in details is one of the
most important problems in modern cosmology, and one of the best
laboratories for this is galactic halos. While CMB and LSS
data probe properties of dark matter on very large scales,
measurements of galactic halos probe small scale properties of dark
matter, such as free-streaming, dark matter self-interactions
\citep{Spergel:1999mh,Dave:2000ar,Hannestad:1999pi,Wandelt:2001ad,%
Colin:2002nk,Burkert:2000di,Firmani:2000ce,Yoshida:2000bx}, dark
matter - baryon interactions \citep{Chen:2002yh,Boehm:2001hm}, or
modified primordial power spectra \citep{Kamionkowski:1999vp,Sigurdsson:2003vy}.

While the method proposed here is not directly applicable to dwarf galaxies where dark matter properties can be probed directly it allows for a very precise determination of the density profiles of massive galaxies. This in turn can shed light on the feedback between baryons and dark matter during the epoch of non-linear structure formation.

We estimate that with future data it will be possible to constrain the
inner density profiles of galactic halos to an accuracy of 1-2\%, much
better than any present measurement. Furthermore our proposed method
mainly probes lens systems at relatively high redshift, $z \gtrsim 1$,
where no other reliable method for measuring galaxy density profiles is
available. 

The estimated precision is based on a sample of roughly 850 multiply
imaged SNe. Even with a significantly smaller set of SNe,
tight constraints could be obtained. This means that properties like
redshift evolution of density profiles, variations in density profiles
with galaxy luminosity, color and many other important 
parameters can be
measured by dividing the lens systems into different categories. This
in turn will provide valuable information on the physics of galaxy
formation.

\section*{Acknowledgments}
The authors wish to thank NORDITA for kind hospitality during the
completion of this work. HD is funded by a post-doctoral fellowship 
from The Research Council of Norway.

\vspace{3cm}


\clearpage

\begin{figure}[htb]
\plotone{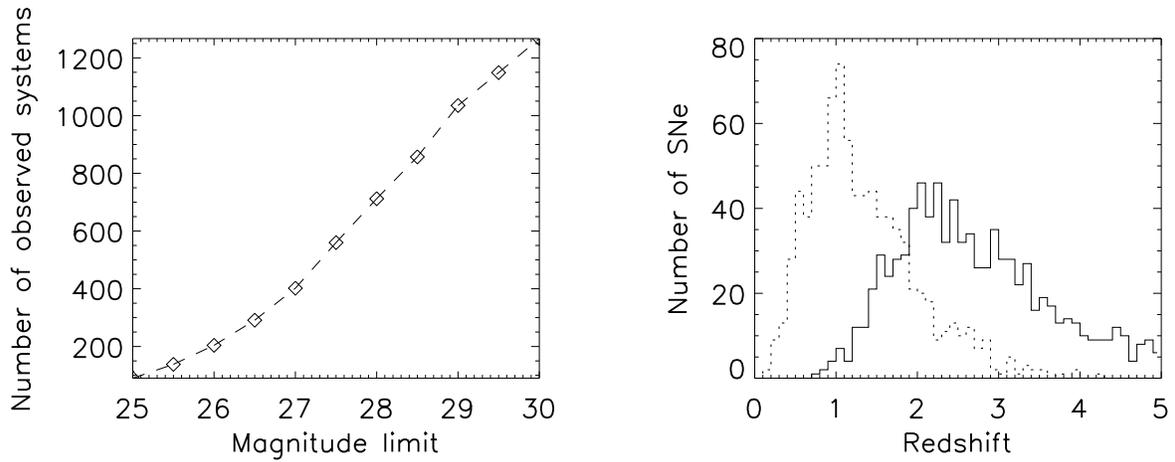}
  \caption{{\em Left panel:} The number of observed multiply imaged
  SNe as a function of magnitude threshold for the dimmest image in
  the I-band or J-band. {\em Right panel:} The solid line shows the
  distribution of source redshifts, the dotted line the lens redshifts
  for the sample of 853 SNe that have either an I-band or J-band
  peak brightness $<28.5$ mag for the dimmest image and
  host galaxy surface brightness $\mu_{\rm I} > 24$ mag at the position of this
  image.}
\label{fig:zdistr}
\end{figure}

\begin{figure}[htb]
\plotone{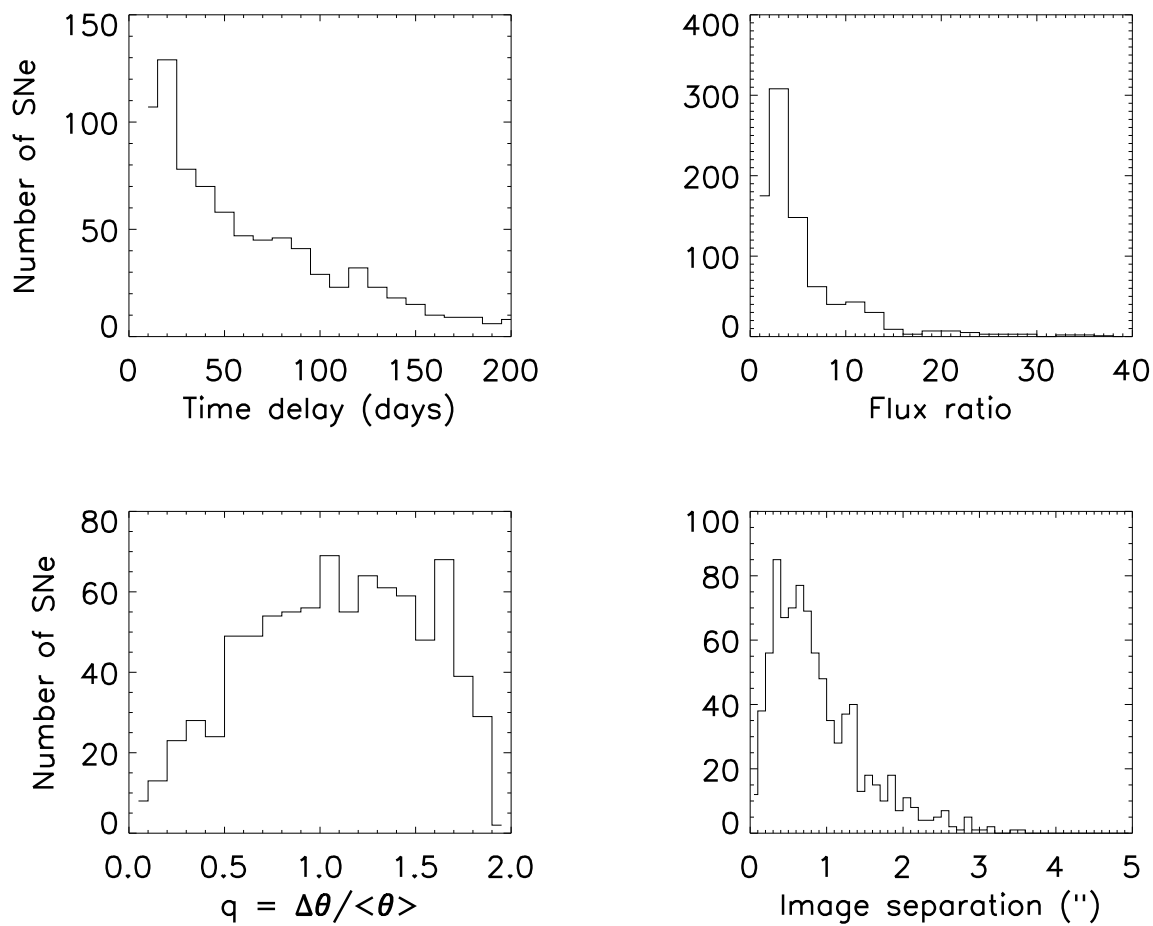}
  \caption{{\em Upper left:} The time delay distribution for 853 SNe
  that have either an I-band or J-band peak brightness $<28.5$ 
  mag for the dimmest image and host galaxy surface
  brightness $\mu_{\rm I} > 24$ mag at this position. {\em Upper right:} The
  distribution of flux ratios for the same sample.  {\em Lower left:}
  The distribution of $q$-values. {\em Lower right:} The image
  separation distribution.}
\label{fig:distr}
\end{figure}

\end{document}